\documentstyle[twocolumn,aps,psfig]{revtex}

\begin{document}
\draft

\twocolumn[\hsize\textwidth\columnwidth\hsize\csname     
@twocolumnfalse\endcsname

\title{Effective Mass of the Four Flux Composite Fermion at $\nu =
1/4$}

\author{W. Pan$^{1,2}$, H.L. Stormer$^{3,4}$, D.C. Tsui$^1$, L.N. Pfeiffer$^3$, 
K.W. Baldwin$^3$, and K.W. West$^3$}
\address{
$^1$Princeton University, Princeton, New Jersey 08544}
\address{
$^2$NHMFL, Tallahassee, Florida 32310}
\address{
$^3$Lucent Technologies, Bell Laboratories, Murray Hill, New Jersey
07974} 
\address{
$^4$Columbia University, New York, New York 10023}

\date{\today}
\maketitle

\begin{abstract}
We have measured the effective mass ($m^*$) of the four flux composite
fermion at Landau level filling factor 
$\nu = 1/4$ ($^4$CF), using the activation energy gaps at the
fractional quantum Hall effect (FQHE) states $\nu$ = 2/7, 3/11, and
4/15 and the temperature dependence of the Shubnikov-de Haas (SdH)
oscillations around $\nu = 1/4$. We find that the energy gaps show a
linear dependence on the effective magnetic field $B_{eff}$ ($\equiv
B-B_{\nu=1/4}$), and from this linear dependence we obtain $m^* = 1.0~m_e$
and a disorder broadening $\Gamma \sim$ 1~K for a sample of density $n
= 0.87 \times 10^{11}$ /cm$^2$. The $m^*$ deduced from the temperature
dependence of the SdH effect shows large differences for $\nu > 1/4$
and $\nu < 1/4$. For $\nu > 1/4$, $m^* \sim 1.0~m_e$. It scales as
$\sqrt{B_{\nu}}$ with the mass derived from the data around $\nu =1/2$
and shows an increase in $m^*$ as $\nu \to 1/4$, resembling the
findings around $\nu =1/2$. For $\nu < 1/4$, $m^*$ increases rapidly
with increasing $B_{eff}$ and can be described by $m^*/m_e = -3.3
+ 5.7 \times B_{eff}$. This anomalous dependence on $B_{eff}$ is
precursory to the formation of the insulating phase at still lower
filling.
\end{abstract}

\vskip2pc]

The composite fermion (CF) model has provided a unified view of the
fractional quantum Hall effect (FQHE) sequences around the
even denominator Landau level filling factors
\cite{sarma:book96,heinonen:book98}. Closely related to Jain's
wavefunction approach \cite{jain:prl89},
Halperin, Lee, and Read \cite{hlr:prb93} 
exploited a singular Chern-Simons
gauge field transformation to explain the electronic transport
properties at, or near, the
even-denominator fillings. In this model, around the Landau level
filling factor $\nu = 1/2\tilde{\phi}$ ($\tilde{\phi}$ is an integer), 
an even number $2\tilde{\phi}$
fictitious magnetic flux quanta are attached to each electron. The so
formed composite particles, still preserving Fermi statistics, are named 
the composite fermions. By so doing,
the strongly interacting two-dimensional electron system
(2DES) in high magnetic ($B$) fields is 
transformed into an equivalent non- or weakly interacting 
CF system experiencing
smaller effective $B$ fields, 
$B_{eff} = B - B_{\nu = 1/2\tilde{\phi}}$ ($B_{\nu =
1/2\tilde{\phi}}$ is the $B$ field at $\nu = 1/2\tilde{\phi}$).
At $\nu = 1/2\tilde{\phi}$, where the attached
flux quanta cancel exactly the external $B$ field (or $B_{eff}$ = 0), 
the CF's form a Fermi sea.
This Fermi system has an effective mass
($m^*$) dependent only on the
electron-electron ($e-e$) interaction, which is proportional to
$e^2/\epsilon l_B$, where $\epsilon$ is the dielectric constant
of GaAs and the magnetic length $l_B = (\hbar c/eB_{\nu})^{1/2}$. 
Thus $m^* \propto \sqrt{B_{\nu}}$. 
When the cancellation is
not exact,
the CF's see $B_{eff}$ and follow
the semiclassical cyclotron orbits. At large $B_{eff}$, cyclotron
orbits of the CF's are quantized and the energy spectrum of the CF's
breaks up into discrete Landau levels. 
The CF Landau level filling factor $p$ is related
to the electron Landau level filling factor $\nu$ through 

\begin{equation}
\nu = \frac{p}{2\tilde{\phi}p \pm 1},
\label{linearT}
\end{equation}
where $p = 1,2,3,\cdots$. 

The above picture was supported by the experiments around $\nu = 1/2$
($\tilde{\phi} = 1$) 
\cite{willett:prl93,kang:prl93,goldman:prl94,smet:prl96,du:cfmass,leadley:prl94,hari:prl94,coleridge:prb95,du:prl95,willett:ss96,yeh:prl99}.
In particular, the activation energy
measurement \cite{du:cfmass}
of the prominent FQHE sequences $\nu = \frac{p}{2p \pm 1}$
showed that the
energy gaps of the FQHE states in each sequence showed
a strikingly linear dependence on $B_{eff}$
emanating from $B_{\nu=1/2}$ (the $B$ field at $\nu = 1/2$),
consistent with the formation of CF's at $\nu
= 1/2$ ($^2$CF's). From this linear dependence an effective
mass, $m^*$,
about one magnitude larger
than the bare 2DES mass was obtained for the CF's
\cite{du:cfmass}.
The Shubnikov-de Haas (SdH)
formalism \cite{du:cfmass,leadley:prl94,hari:prl94,coleridge:prb95} 
was also adopted to determine $m^*$ and the scattering
time of $^2$CF's. It was found from this analysis that $m^*$ was
nearly constant for large positive and negative $B_{eff}$,
showed a $B_{eff}$ dependent
enhancement for smaller $B_{eff}$, and divergent as $B_{eff} \to 0$.

The focus of all earlier experiments was on the $^2$CF's 
around $\nu = 1/2$ or
around $\nu = 3/2$ 
\cite{du:prl95}.
Other even-denominator filling factors, such as $\nu = 1/4$ 
($\tilde{\phi}$ = 2),
was not fully addressed.
In particular, will the CF's form at $\nu = 1/4$? It is believed that
Wigner crystal is the preferred ground state for $\nu < 1/5$. 
If the CF's do form at
$\nu = 1/4$ ($^4$CF's), how do they 
compare to the $^2$CF's?  So far, only a few preliminary experimental 
results about $\nu = 1/4$
exist in the literature \cite{du:cfmass,leadley:prl94,willett:ss96}. 
The recent study carried out by Yeh $et~al.$ \cite{yeh:prl99} 
concentrated on the $^4$CF's at $\nu = 3/4$,
the particle-hole conjugate of $\nu = 1/4$.
It should be apparent that this lack of 
experiments on the $^4$CF's is mainly 
due to the fact that such an experiment is far more demanding.
To resolve
the FQHE sequences $\nu = \frac{p}{4p \pm 1}$ around $\nu = 1/4$
requires much higher quality samples and higher $B$ fields.

We wish to report in this paper a systematic study of the $^4$CF's
at $\nu = 1/4$. We find that 
the energy gaps of
the prominent FQHE states above $\nu = 1/4$, $i.e.$, at $\nu$ = 
2/7, 3/11, and 4/15, show a
linear dependence on $B_{eff}$ ($\equiv B - B_{\nu = 1/4}$). From
this linear dependence we obtain an effective mass $m^*/m_e = 1.0$,
where $m_e$ is the bare electron mass, 
and a disorder broadening $\Gamma \sim$ 1 K for a sample of
density $n = 0.87 \times 10^{11}$ /cm$^2$.
We have also studied
the temperature dependence of the $\nu = \frac{p}{4p \pm 1}$
FQHE series as that of the SdH effect of the $^4$CF's and
determined $m^*$ as a function of $B_{eff}$ on both sides of 
$\nu = 1/4$.
The $m^*$ deduced from the data taken for
$B_{eff} < 0$ and from those taken for $B_{eff} > 0$ 
differs considerably. 
For large negative $B_{eff}$ ($\leq -1.0$~T), $m^*$ is nearly constant
($m^*/m_e \sim 1.0$), in agreement with that 
from the activation measurement. It scales as
$\sqrt{B_{\nu}}$ with the mass derived from the data 
at lower B field around $\nu = 1/2$. An
increase in $m^*$ is observed as $B_{eff} \to 0$, resembling 
the findings around $\nu = 1/2$.
However, for $B_{eff} > 0$, $m^*$
increases rapidly with $B_{eff}$ and
can be described by $m^*/m_e = -3.3 + 5.7
\times B_{eff}$ ($B_{eff}$ in Tesla). We believe that this 
anomalous linear dependence on 
$B_{eff}$ is related to the formation of
the insulating phase
at still higher $B$, or smaller $\nu$.

We have investigated three samples (A, B, and C in Table I) 
from three different wafers. 
The samples are GaAs/AlGaAs heterostructures of electron densities 
$n$ = (0.87, 0.85, 0.65) $\times~10^{11}$ /cm$^2$
and mobilities $\mu$ = (10.5, 8.2, 6.6)
$\times~10^6$ cm$^2$/Vs, respectively,
after illumination by a red light-emitting diode (LED) at 4.2 K.
The size of each sample is about 5 mm $\times$ 5 mm 
with eight indium contacts placed symmetrically around
the edges, four at the corners and four in the centers of the
four edges. A standard low-frequency ($\sim$ 5Hz)
lock-in technique was employed to measure the transport coefficients,
using an excitation current varying from 0.1~nA to 1~nA.
A calibrated $RuO_2$ thermometer of
known corrections for magnetoresistance was
mounted next to the sample for the temperature ($T$) measurement. 
The experiments were performed in a top-loading 
dilution refrigerator with $T$ down to 
20~mK and $B$ up to 18~T.

Fig.~1 shows the diagonal resistance $R_{xx}$ vs. $B$ 
around $\nu=1/4$ for sample A.
The exceptional sample quality is manifested by the 
appearance of the higher order FQHE states at
$\nu =$ 5/19, 4/17 and a very low resistance at $\nu =
1/5$. $R_{xx}$ has been  measured at many temperatures, and here we
show two 
different temperature traces taken at $T$ = 25 and 150 mK. 
The FQHE states at $\nu$ = 2/7, 3/11, and 4/15 above $\nu = 1/4$ and
at $\nu = 2/9$ below $\nu = 1/4$ exhibit the typical quantum Hall
liquid characteristic --- the resistance minima 
decrease with decreasing $T$.
The activation gap $\Delta$ is obtained
from the $T$ dependence of $R_{xx}$ 
through $R_{xx} \propto$
exp(-$\Delta/2k_BT$). In the inset, the activation
gaps at 2/7, 3/11, and 4/15
are plotted against $B_{eff} = B - B_{\nu = 1/4}$. 
They show a linear dependence on $B_{eff}$
with a negative intercept $\Gamma$ at $B_{eff} = 0$.
This linear dependence, as expected for the $^4$CF's at
$\nu = 1/4$,
allows us to obtain an effective mass, $m^*/m_e = 1.0$,
from the slope of the line
via the equation $\Delta =
\hbar eB_{eff}/m^*c - \Gamma$ \cite{hlr:prb93,du:cfmass}.
The negative intercept at
$B_{eff} = 0$ gives the disorder broadening
of the CF Landau level, $\Gamma \sim$ 1 K.

We have also studied the $T$ dependence of the $\nu = \frac{p}{4p
\pm 1}$ FQHE series as that of the SdH effect of
the $^4$CF's and deduced $m^*$
on both sides of $\nu = 1/4$
\cite{du:cfmass,leadley:prl94,hari:prl94,coleridge:prb95}.
The amplitude of the SdH oscillations ($\Delta R$) is 
obtained following the standard method \cite{du:cfmass}, and
is fitted  by a nonlinear least-square technique 
according to the following equation

\begin{equation}
\frac{\Delta R}{R_{0}} \propto \frac{\xi}{sinh(\xi)},
\label{linearT}
\end{equation}
where $\xi=2\pi^2k_{B}T/\hbar\omega_{c}^*=2\pi^2k_{B}Tm^*c/e\hbar B_{eff}$,
$\omega_{c}^*$ is the cyclotron frequency, and $m^*$ is the effective
mass. $R_{0}$ is the diagonal resistance at
$B_{eff} = 0$. All other symbols have their usual meanings. 
In the fitting, it has
been assumed that the relaxation time $\tau$ of the $^4$CF's is
temperature independent. 

Since $m^* \propto \sqrt{B_{\nu}}$, we
introduce a normalized mass,
$m^*_{nor} = m^*/\sqrt{B_{\nu}}$ to take into account of the sample
density differences. Here $m^*$ is in units of $m_e$ 
and $B_{\nu}$ is the magnetic field at which $B_{eff} = 0$, in units of T.
In Fig.~2, $m^*_{nor}$ for the CF's at $\nu = 1/4$
is plotted as a function of
$B_{eff}$ for all three samples: solid squares for sample A, 
open circles for B, and open diamonds for C.
We also include the $m^*_{nor}$ for the $^4$CF's
at $\nu = 3/4$ from samples D, E, and F
of Yeh {\it et~al.} \cite{yeh:prl99}
(noting that $B_{eff} = 
-3 \times (B-B_{\nu = 3/4})$ for the $^4$CF's around $\nu = 3/4$).
In Table I we summarize the sample parameters
for all six samples.
Despite the large differences in $n$ and $\mu$, the normalized $^4$CF
mass behaves very similarly:
$m^*_{nor} \sim 0.26$
for $B_{eff} \leq -1.0$~T,
increases to $\sim 0.33$ at $B_{eff} = -0.7$~T, and tends to
diverge as $B_{eff} \to 0$, similar to
the findings for $^2$CF's.
For $B_{eff} > 0$, 
$m^*_{nor}$ is anomalously large. It shows a linear 
dependence on $B_{eff}$ and can be fitted
to $m^*_{nor} = -0.88 + 1.5 \times B_{eff}$, 
independent of $n$ and $\mu$ and whether  
it is measured around $\nu= 1/4$ or $\nu = 3/4$.

It is clear from Fig.~1 that the $\nu = \frac{p}{4p \pm 1}$ FQHE
features in $R_{xx}$ vs. $B$ show a large asymmetry across $\nu = 1/4$,
and for $\nu < 1/4$, a strong $T$ dependent background is
observed  to  increase steeply with increasing $B$.
The huge $R_{xx}$ peak ($\sim 4$ M$\Omega$) 
between $\nu = 2/9$ and $\nu = 1/5$ is
due to the reentrant insulating behavior attributed to the formation
of a pinned Wigner solid
\cite{reentrant}. Hence, it is not surprising that the $^4$CF mass
deduced from the $T$ dependence of the $\nu = \frac{p}{4p \pm 1}$ FQHE
features is anomalous, and it is natural to
associate the observed asymmetry in $m^*$ vs. $B_{eff}$
with the formation of
the insulating phase at still lower filling factors. It should also be
noted that anomalously large $^4$CF $m^*$ was previously deduced from the
$\nu = 4/5$ FQHE data, and it was believed to be a consequence of the
influence of hole localization of the electron-hole symmetric state at
$\nu = 4/5$. In any case, the $m^*$ in Table I is the $^4$CF mass
obtained at  
large negative $B_{eff}$
around $\nu = 1/4$ and 3/4. 

We have also calculated $m^*$ at $\nu = 2/7$ for sample A following the 
mean field theory \cite{hlr:prb93}, which relates $m^*$ to the Coulomb
energy $e^2/\epsilon l_B$ via
$\hbar eB_{eff}/m^*c = g^{\nu} e^2/\epsilon l_B$.
Here $g^{\nu}$ is a constant at $\nu$
\cite{fano:prb86,fqhe:book,morf:prb89,nick:prb95} and
we used $g^{(\nu=2/7)} = 0.024$ from Ref.~20.
$m^*$ was found, after corrected for
the finite thickness of the 2DES \cite{zhang:prb86}, to be 0.9 $m_e$, very close
to the measured value of 1.0 $m_e$. 

We now turn to the comparison between the $^4$CF's and the $^2$CF's.
For this purpose, we measured
the SdH $m^*$ of the $^2$CF's of sample A, and plotted them 
as the solid circles in Fig.~3.
The solid squares are the $^4$CF data 
for $B_{eff} < 0$ of the same 
sample. The $^4$CF data for  
$B_{eff} > 0$ are not included.
We also include the $^2$CF data from  
Refs.~9 and 12; they are
represented by the open symbols --- squares, circles, dot-centered
up-triangles, and cross-centered down-triangles.
Two important results emerge from Fig.~3. First,
the $m^*_{nor}$'s for $^4$CF and $^2$CF from the same sample are of the same
magnitude within
our experimental error and show a similar
dependence on $B_{eff}$. 
So far we are not aware of any theories
predicting a same value of $m^*$ for the $^2$CF's and $^4$CF's.
On the contrary, if we regard the $\nu=1/4$ state as 
the $\nu = 1/2$ state of CF's emanating from $\nu = 1/2$, 
we would have expected to see different $m^*$'s. 
Second, the $^2$CF mass
from different samples with different $n$ and $\mu$, varying from $n =
0.87 \times 10^{11}$ /cm$^2$ to $2.3 \times 10^{11}$ /cm$^2$ and $\mu
= 2.5 \times 10^6$ cm$^2$/Vs to $12.8 \times 10^6$ cm$^2$/Vs,
also shows a similar 
dependence on $B_{eff}$. The large spread in magnitude is to be
expected in view of the large difference in disorder in the different
samples as seen in the spread in their mobilities.

In summary, we have measured the $^4$CF effective mass using the activation
energy gaps at the FQHE states $\nu$ = 2/7, 3/11, and 4/15, and the
temperature dependence of the SdH oscillations around $\nu = 1/4$. We
find that the energy gaps show a linear dependence on $B_{eff}$, and
from this linear dependence we obtain $m^* = 1.0~m_e$ and a disorder
broadening $\Gamma \sim 1$~K. The $m^*$ deduced from the $T$
dependence of the SdH effect shows an asymmetry for the negative and
positive $B_{eff}$. For $B_{eff} < 0$, $m^* \sim 1.0~m_e$ 
for large negative $B_{eff}$ ($\leq -1.0$~T), shows a
$B_{eff}$ dependent enhancement for smaller $B_{eff}$, and divergent
as $B_{eff} \to 0$. It scales as $\sqrt{B_{\nu}}$ with the mass
derived from the data around $\nu$ = 1/2 and 3/4. For $B_{eff} > 0$,
$m^*$ is anomalously large and shows a linear dependence on $B_{eff}$.
This behavior is
believed to be related to the formation of Wigner crystal phase at
still lower filling factor.

We thank E. Palm and T. Murphy
for the experimental assistance, and A.S. Yeh, R.R. Du, and 
N.E. Bonesteel for discussions. 
A portion of this work was performed at the National 
High Magnetic Field Laboratory which is supported by NSF Cooperative
Agreement No. DMR-9527035 and by the State of Florida. D.C.T. and W.P.
are supported by the DOE and the NSF.

\begin{table}
\vspace{8.2cm}
\caption{
The density $n$ and the mobility $\mu$ for samples (A, B, C) and (D, E,
F)$^{15}$, and the effective mass $m^*$ for $^4$CF measured
around $\nu = 1/4$ and $\nu = 3/4$. 
$m^*$ is in units of the bare electron mass.
}
\begin{tabular}{ccccc}
sample &
{\rm $n$ (10$^{11}$ /cm$^2$)} &
{\rm $\mu$ (10$^6$ cm$^2$/Vs)} &
{\rm $m^*$} &
{\rm $m^*_{nor}$} \\ 
\tableline
A (03-14-97-1A) & 0.87 & 10.5 & 1.0 & 0.27 \\
B (03-27-97-1A) & 0.85 & 8.2 & 1.0 & 0.27 \\
C (03-27-97-2A) & 0.65 & 6.6 & 0.90 & 0.29 \\
\tableline
D (02-03-97-1A) & 0.86 & 11.0 & 0.58 & 0.26 \\
E (10-19-91-1A) & 1.13 & 6.8 & & \\
F (12-06-91-1A) & 2.26 & 13.0 & 0.90 & 0.25 \\
\end{tabular}
\end{table}

\begin{figure}
\centerline{\psfig{figure=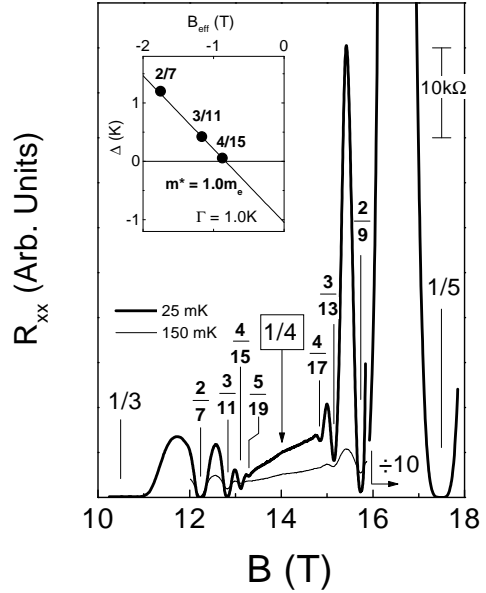,width=7.5cm,angle=0}}
\caption{
$R_{xx}$ vs. $B$ around $\nu = 1/4$ for sample A.
The vertical lines mark the positions of
filling factor $\nu$.
Two traces at 25 mK
(thick line) and 150 mK (thin line) are shown to illustrate the
temperature dependence of $R_{xx}$. The inset shows the plot of the
activation gap $\Delta$ vs. $B_{eff}$ at $\nu$ = 2/7, 3/11, and 4/15. 
The straight line is a linear fit to the experimental data. 
$m^*$ is the effective mass of the $^4$CF's obtained from the activation
measurement. $\Gamma$ is the disorder broadening.}
\end{figure}

\begin{figure}
\centerline{\psfig{figure=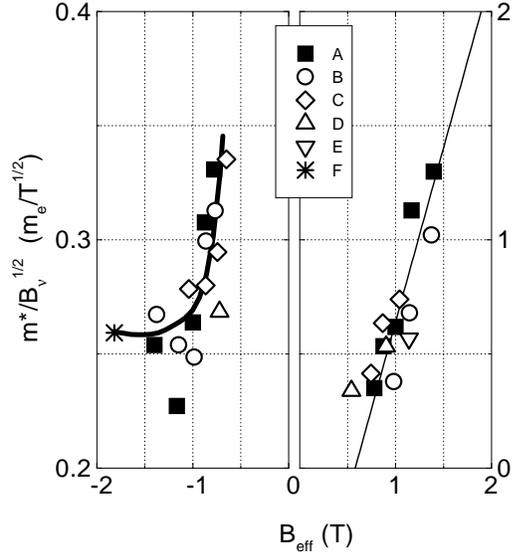,width=7.5cm,angle=0}}
\caption{
$m^*_{nor}$ vs. $B_{eff}$ around $\nu = 1/4$ 
for sample A (solid squares), B (open circles), and C (open diamonds).
The results around $\nu = 3/4$ obtained by Yeh $et~al.^{15}$ for 
sample D (open up-triangles), E (open
down-triangle), and F (aster) are also included. The gray line is guide
to the eye and the straight line is a linear fit to the data. $m^*$ is
in units of $m_e$. $B_{\nu}$ is the $B$ field where the CF's form, in
units of T.} 
\end{figure}

\begin{figure}
\centerline{\psfig{figure=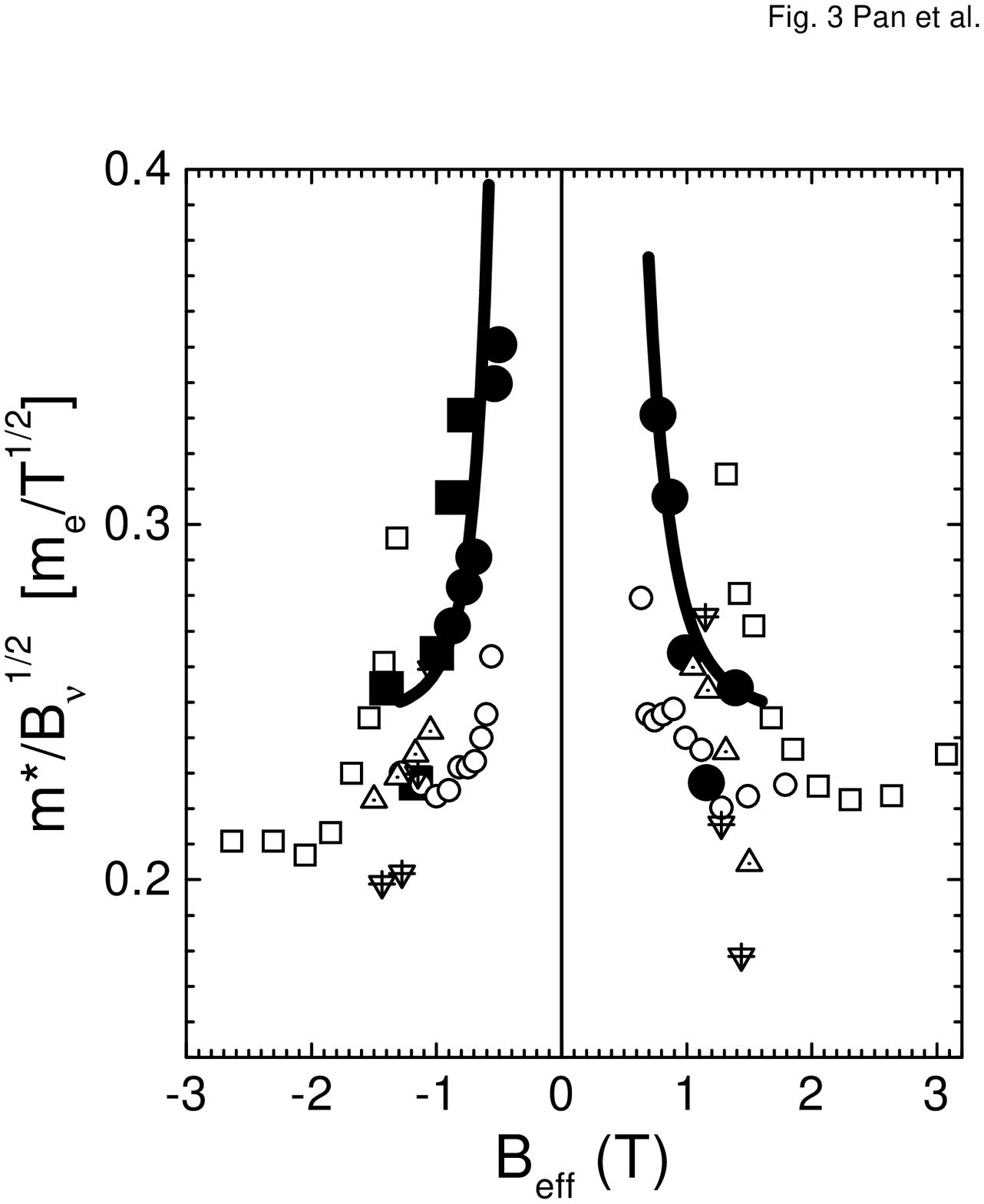,width=7.5cm,angle=0}}
\caption{
Comparison of $m^*_{nor}$ of the $^4$CF's and the $^2$CF's. 
The solid squares and solid circles represent $m^*_{nor}$ for $^4$CF 
and $^2$CF respectively of sample A. The gray lines are guides to the
eye. The open symbols (\protect{\large $\Box$}, $\bigcirc$, $\triangle$ \hskip -0.35cm $\cdot$~,
and $\bigtriangledown$ \hskip -0.4cm $+$)
represent the $^2$CF data obtained by Du {\it et al.}$^9$ and
Coleridge {\it et al.}$^{12}$ for their samples of densities (2.3, 1.13,
1.27, and 1.39) $\times 10^{11}$ /cm$^2$ and mobilities (12.8, 6.8,
3.5, and 2.5) $\times 10^6$ cm$^2$/Vs, respectively.
$m^*$ is in units of $m_e$ and
$B_{\nu}$ in T.} 
\end{figure}


\begin{references}

\bibitem{sarma:book96}
	{\it Perspectives in Quantum Hall
Effects}\, edited by S. Das Sarma and A. Pinczuk (Wiely, New York,
1996).

\bibitem{heinonen:book98}
	{\it Composite Fermions: A Unified View of the Quantum Hall
Regime}\, edited by O. Heinonen (World Scientific, Singapor, 1998).

\bibitem{jain:prl89}
	J.K. Jain, Phys.\ Rev.\ Lett. {\bf 63}, 199 (1989); Phys.\ Rev.\
B {\bf 41}, 7653 (1990); Science {\bf 266}, 1199 (1994).

\bibitem{hlr:prb93}
	B.I. Halperin, P.A. Lee, and N.Read,
	Phys.\ Rev.\ B {\bf 47}, 7312 (1993).

\bibitem{willett:prl93}
	R.L. Willett, R.R. Ruel, K.W. West, and L.N. Pfeiffer,
	Phys.\ Rev.\ Lett. {\bf 71}, 3846 (1993).

\bibitem{kang:prl93}
	W. Kang, H.L. Stormer, L.N. Pfeiffer, K.W. Baldwin, and K.W. West,
	Phys.\ Rev.\ Lett. {\bf 71}, 3850 (1993).

\bibitem{goldman:prl94}
	V.J. Goldman, B. Su, and J.K. Jain,
	Phys.\ Rev.\ Lett. {\bf 72}, 2065 (1994).

\bibitem{smet:prl96}
	J.H. Smet, D. Weiss, R.H. Lutjering, K. von Klitzing, R.
Fleishmann, R. Ketmerick, T. Geisel, and G. Weimann,
	Phys.\ Rev.\ Lett. {\bf 77}, 2272 (1996).

\bibitem{du:cfmass}
	R.R. Du, H.L. Stormer, D.C. Tsui, L.N. Pfeiffer, and K.W.
West,
	Phys.\ Rev.\ Lett. {\bf 70}, 2944 (1993); {\it ibid.}
{\bf 73}, 3274 (1994); Solid\ State\ Communications {\bf 90}, 71
(1994).

\bibitem{leadley:prl94}
	D.R. Leadley, R.J. Nicholas, C.T. Foxon, and J.J. Harris,
	Phys.\ Rev.\ Lett. {\bf 72}, 1906 (1994).

\bibitem{hari:prl94}
	H.C. Manoharan, M. Shayegan, and S.J. Klepper,
	Phys.\ Rev.\ Lett. {\bf 73}, 3270 (1994).

\bibitem{coleridge:prb95}
	P.T. Coleridge, Z.W. Wasilewski, P. Zawadski, A.S. Sachrajda,
and H.A. Carmona,
	Phys.\ Rev.\ B {\bf 52}, 11603 (1995).

\bibitem{du:prl95}
        R.R. Du, A.S. Yeh, H.L. Stormer, D.C. Tsui, L.N. Pfeiffer, and K.W.
West,
        Phys.\ Rev.\ Lett. {\bf 75}, 3926 (1995).

\bibitem{willett:ss96}
	R.L. Willett and L.N. Pfeiffer,
	Surface Science {\bf 361/362}, 38 (1996).

\bibitem{yeh:prl99}
	A.S. Yeh, H.L. Stormer, D.C. Tsui, L.N. Pfeiffer, K.W. Baldwin,
and K.W. West,
	Phys.\ Rev.\ Lett. {\bf 82}, 592 (1999).

\bibitem{reentrant}
	H.W. Jiang, R.L. Willett, H.L. Stormer, D.C. Tsui, L.N.
Pfeiffer, and K.W. West,
	Phys.\ Rev.\ Lett. {\bf 65}, 633 (1990);
	R.R. Du, D.C. Tsui, H.L. Stormer, L.N. Pfeiffer, and K.W.
West,
	Solid\ State\ Communications {\bf 99}, 755 (1996). 

\bibitem{fano:prb86}
        G. Fano, F. Ortolani, and E. Colombo,
        Phys.\ Rev.\ B {\bf 34}, 2670 (1986).

\bibitem{fqhe:book}
	T. Chakraborty and P. Pietil$\ddot{a}$inen, {\it The
Fractional Quantum Hall Effect}, Springer Series in solid State
Science {\bf 85} (Springer-Velag, New York, 1988).

\bibitem{morf:prb89}
	N. d'Ambrumenil and R. Morf,
	Phys.\ Rev.\ B {\bf 40}, 6108 (1989).

\bibitem{nick:prb95}
	N.E. Bonesteel,
	Phys.\ Rev.\ B {\bf 51}, 9917 (1995).

\bibitem{zhang:prb86}
	F.C. Zhang and S. Das Sarma, 
	Phys.\ Rev.\ B {\bf 33}, 2903 (1986).

%
\end{references}
\end{document}